\documentclass[conference]{IEEEtran}
\usepackage[T1]{fontenc}
\usepackage{cite}
\usepackage{amsmath,amssymb}
\usepackage{algorithm}
\usepackage[noend]{algpseudocode}
\usepackage{graphicx}
\usepackage{booktabs}
\usepackage{multirow}
\usepackage{threeparttable}
\usepackage{tabularx}
\usepackage{array}
\usepackage{balance}
\usepackage{url}
\usepackage{microtype}
\usepackage[dvipsnames]{xcolor}
\usepackage{tikz}
\usepackage{pgfplots}
\usepackage{pgfplotstable}
\usetikzlibrary{arrows.meta,positioning,calc,fit,backgrounds,patterns}
\usepgfplotslibrary{groupplots,statistics}
\pgfplotsset{compat=1.17}

\renewcommand{\texttt}[1]{{\ttfamily\bfseries #1}}

\newcommand{\vect}[1]{\boldsymbol{#1}}

\newcommand{\cB}{\mathcal{B}}
\newcommand{\cC}{\mathcal{C}}
\newcommand{\cF}{\mathcal{F}}
\newcommand{\cG}{\mathcal{G}}
\newcommand{\cL}{\mathcal{L}}
\newcommand{\cT}{\mathcal{T}}
\newcommand{\firstK}{\operatorname{first}_K}
\newcommand{\figref}[1]{Fig.~\ref{#1}}
\newcommand{\Figref}[1]{Figure~\ref{#1}}
\newcommand{\tabref}[1]{Table~\ref{#1}}
\newcommand{\Tabref}[1]{Table~\ref{#1}}
\newcommand{\secref}[1]{Section~\ref{#1}}
\newcommand{\Secref}[1]{Section~\ref{#1}}
\newcommand{\eqnref}[1]{(\ref{#1})}
\newcommand{\Algref}[1]{Algorithm~\ref{#1}}

\makeatletter
\g@addto@macro\normalsize{%
  \setlength\abovedisplayskip{1.5pt plus 0.5pt minus 1pt}%
  \setlength\belowdisplayskip{1.5pt plus 0.5pt minus 1pt}%
  \setlength\abovedisplayshortskip{1pt plus 0.5pt minus 1pt}%
  \setlength\belowdisplayshortskip{1pt plus 0.5pt minus 1pt}%
}
\renewcommand\section{\@startsection{section}{1}{\z@}{1.15ex plus 0.05ex minus 0.2ex}{0.45ex plus 0.05ex minus 0ex}{\normalfont\normalsize\centering\scshape}}
\renewcommand\subsection{\@startsection{subsection}{2}{\z@}{1.1ex plus 0.05ex minus 0.2ex}{0.45ex plus 0.05ex minus 0ex}{\normalfont\normalsize\itshape}}
\makeatother
\begin{document}

\title{Feasibility-Aware Security-Constrained Unit Commitment via Hybrid Soft Actor-Critic with Quantum-Sampled Features}

\author{\IEEEauthorblockN{George Dimas, Amin Masoumi, and Mert Korkali}
\IEEEauthorblockA{\textit{Department of Electrical Engineering and Computer Science}\\
\textit{University of Missouri}\\
Columbia, MO 65211 USA\\
E-mail: {\{gadtbg, am4n5, korkalim\}@missouri.edu}}
}

\maketitle
\bstctlcite{IEEEexample:BSTcontrol}

\begin{abstract}
Security-constrained unit commitment (SCUC) couples binary commitment, economic dispatch, reserves, and network security over a multiperiod horizon, making an exact solution computationally expensive for realistic system sizes. This paper proposes a three-layer hybrid framework in which a Bernoulli hybrid soft actor-critic (HSAC) policy proposes hourly commitments, a quantum-sampled auxiliary channel augments the state, and a native SCUC mixed-integer linear program recovers dispatch and security variables after only a limited subset of commitment binaries is enforced. The method is therefore solver-compatible rather than an end-to-end replacement for exact optimization. We formalize the SCUC-to-reinforcement-learning interface, derive the temporal coverage induced by the fixed cap, and evaluate the \mbox{14-,} \mbox{57-,} and 118-bus benchmark cases. The results show stable, low-cost recovery in the 14-bus case, where the best recovered schedule attains the full-horizon optimum; a very low screen-rejection rate in the 57-bus case; and a clear coverage bottleneck in the 118-bus case once the enforcement cap no longer spans a complete commitment period. The study, therefore, identifies the amount of useful commitment information that reaches the recovery model, under an exploratory Bernoulli actor and a small enforcement cap, as the dominant limitation that governs scalability.
\end{abstract}


\section{Introduction}
\label{sec:intro}
Security-constrained unit commitment (SCUC) is a central optimization problem in power-system operations and electricity markets, as it must coordinate binary generator commitments, continuous dispatch, reserve requirements, and transmission-security constraints over a rolling horizon. Even when the network is linearized with a direct-current (DC) approximation, the resulting mixed-integer linear program (MILP) remains computationally demanding at realistic system scales \cite{chen2023scuc,yang2021ml}. The practical need is considerable; day-ahead markets must clear SCUC for thousands of units within a window of a few hours, reliability re-commitment during the operating day leaves only minutes, and renewable uncertainty multiplies the number of solutions required within each window \cite{chen2023scuc}, so methods that reduce this burden without weakening solver feasibility guarantees have direct operational value.
Recent research has explored learning-assisted SCUC along several complementary lines. Xavier \emph{et al.} \cite{xavier2021learning} learn warm starts and constraint screening for market-scale SCUC, whereas Pineda \emph{et al.} \cite{pineda2022low} caution that learned commitment predictions degrade without a feasibility-restoration mechanism; subsequent work restricts the integer search space through explanation-guided variable reduction \cite{dai2025reduction}, fuzzy feasibility guarantees \cite{venkatesh2025feasibility}, structure-aware masking with solver-preserving guarantees \cite{wang2026structure}, and successive fixing driven by first-order methods \cite{xiong2025successive}. Reinforcement learning (RL) has likewise been applied to unit commitment (UC); de~Mars and O'Sullivan combine policy learning with guided tree search \cite{demars2021applying,demars2022astar}, Qin \emph{et al.} \cite{qin2023ensemble} assist an ensemble of RL agents with exact optimization, Sayed \emph{et al.} \cite{sayed2024acuc} couple RL with convex programming for AC-feasible UC, and further variants address wind uncertainty, adaptive horizons, expert knowledge, and graph-structured policies \cite{xu2024wind,yan2024lookahead,liang2024expert,lu2026graph}. In parallel, quantum and hybrid quantum-classical studies have examined the quantum approximate optimization algorithm (QAOA) \cite{koretsky2021qaoa}, quantum annealing \cite{hong2025qa}, surrogate Lagrangian decomposition \cite{feng2023surrogate}, quantum-search and quantum-neural formulations \cite{zheng2024fastquantum,liu2025exact,wei2026qrl}, and distributed or hybrid quantum-classical recovery schemes \cite{aboumrad2025hybrid,hasanzadeh2025d2uc,barrass2025leveraging}; a recent survey appears in \cite{hasanzadeh2026survey}. These directions are valuable, but two practical gaps remain.
First, end-to-end RL policies often require a repair layer because enforcing exact UC feasibility directly in the policy space is difficult, especially when intertemporal and network constraints are active \cite{pineda2022low,sayed2024acuc}. Second, many quantum UC studies rely on quadratic unconstrained binary optimization (QUBO) reformulations or surrogate decompositions, whose scalability remains limited by near-term hardware and embedding constraints \cite{koretsky2021qaoa,hong2025qa,aboumrad2025hybrid,hasanzadeh2026survey}. These observations motivate a more conservative integration strategy, namely to retain the native SCUC solver and learn only a compact subset of commitment decisions that meaningfully shrinks the binary search space.
This paper adopts exactly that strategy through a \textit{three-layer hybrid framework}.\footnote{The three layers are policy generation, quantum feature sampling, and classical SCUC recovery.} The proposed method couples a Bernoulli hybrid soft actor-critic (HSAC) policy, a quantum-sampled state-augmentation channel, a capacity screen, and a warm-started SCUC recovery model that enforces only a limited subset of policy-proposed commitment binaries. Commitment tuples are appended in time order, and only the first $K$ are imposed as equality constraints in the recovery model. The main contributions are threefold. First, we formulate a solver-compatible RL-to-SCUC interface in which policy outputs restrict only part of the binary space, while the SCUC model retains responsibility for dispatch, reserves, and security feasibility. Second, we tailor soft actor-critic (SAC) to multi-binary UC actions for entropy-regularized learning with a mixed-integer recovery layer. Third, we show that the current chronological enforcement rule induces a measurable coverage ratio that strongly influences scalability.\footnote{Code, training scripts, and result traces to reproduce all results are openly available at \url{https://github.com/GeorgeDimas123/QHSAC-Unit-Commitment} \cite{hsacrepo2026}.}

The remainder of this paper is organized as follows. \Secref{sec:formulation} formulates SCUC and the sequential decision mapping; \Secref{sec:method} describes the proposed HSAC-SCUC method; \Secref{sec:setup} summarizes the implementation and experimental setup; \Secref{sec:results} presents the results and discussion; and \Secref{sec:conclusion} concludes the paper.

\section{SCUC Formulation and Sequential Decision Mapping}
\label{sec:formulation}
This section presents the SCUC model that the recovery layer solves and then maps partial-commitment enforcement to a sequential decision process.
\subsection{SCUC Model}
Let $\cG$, $\cB$, $\cL$, $\cC$, and $\cT=\{1,\ldots,T\}$ denote the sets of thermal units, buses, transmission lines, contingencies, and time periods. We use $N_g=|\cG|$, $N_b=|\cB|$, and $N_l=|\cL|$. For each Bus $b\in\cB$, let $\cG_b\subseteq\cG$ be the set of units connected to $b$, and let $\delta(b)\subseteq\cL$ denote the incident lines. For each Line $l\in\cL$, let $i(l)$ and $j(l)$ denote its sending- and receiving-end buses. For each Generator $g\in\cG$ and Period $t\in\cT$, the binary variables $u_{g,t}$, $v_{g,t}$, and $w_{g,t}$ denote the on/off, startup, and shutdown decisions, respectively. The continuous variables $p_{g,t}$, $r_{g,t}$, $f_{l,t}$, $\theta_{b,t}$, and $\ell_{b,t}^{\mathrm{ls}}$ denote active-power dispatch, reserve, line flow, phase angle, and involuntary load shedding, respectively. The demand at Bus $b$ and Time $t$ is $d_{b,t}$; the total system demand is $D_t=\sum_{b\in\cB} d_{b,t}$; the reserve requirement is $R_t^{\mathrm{req}}$; and $C_g(\cdot)$ denotes the production-cost function of Unit $g$, i.e., a convex or piecewise-linear approximation of dispatch cost.
The total operating cost $Z$ collects no-load, startup, shutdown, dispatch, and load-shedding costs, viz.,
\begin{align}
Z &= \sum_{t\in\cT}\sum_{g\in\cG} \Big(c_g^{\mathrm{nl}}u_{g,t}+c_g^{\mathrm{su}}v_{g,t}+c_g^{\mathrm{sd}}w_{g,t}+C_g(p_{g,t})\Big) \nonumber\\
&\quad + \sum_{t\in\cT}\sum_{b\in\cB} c^{\mathrm{ls}}\ell_{b,t}^{\mathrm{ls}},
\label{eq:obj}
\end{align}
where $c_g^{\mathrm{nl}}$, $c_g^{\mathrm{su}}$, and $c_g^{\mathrm{sd}}$ are the no-load, startup, and shutdown cost coefficients of Unit $g$, and $c^{\mathrm{ls}}$ is the load-shedding penalty price. The SCUC problem minimizes $Z$ subject to the representative constraint set used throughout the paper, viz.,
\begin{subequations}\label{eq:scuc}
\begin{align}
u_{g,t}-u_{g,t-1} &= v_{g,t}-w_{g,t}, && \forall g,t, \label{eq:logic}\\
\sum_{\tau=t}^{t+UT_g-1} u_{g,\tau} &\ge UT_g v_{g,t}, && \forall g,t, \label{eq:uptime}\\
\sum_{\tau=t}^{t+DT_g-1} (1-u_{g,\tau}) &\ge DT_g w_{g,t}, && \forall g,t, \label{eq:dntime}\\
\underline P_g u_{g,t} \le p_{g,t} &\le \overline P_g u_{g,t}, && \forall g,t, \label{eq:pbound}\\
-RD_g \le p_{g,t}-p_{g,t-1} &\le RU_g, && \forall g,t, \label{eq:ramp}\\
\sum_{g\in\cG_b} p_{g,t} - d_{b,t} + \ell_{b,t}^{\mathrm{ls}} &= \sum_{l\in\delta(b)} f_{l,t}, && \forall b,t, \label{eq:balance}\\
f_{l,t} &= B_l\big(\theta_{i(l),t}-\theta_{j(l),t}\big), && \forall l,t, \label{eq:dcflow}\\
|f_{l,t}| &\le F_{l}^{\max}, && \forall l,t, \label{eq:flowlimit}\\
p_{g,t}+r_{g,t} &\le \overline P_g u_{g,t}, && \forall g,t, \label{eq:headroom}\\
\sum_{g\in\cG} r_{g,t} &\ge R_t^{\mathrm{req}}, && \forall t, \label{eq:reserve}\\
|f_{l,t}^{(c)}| &\le F_{l,c}^{\max}, && \forall l,c,t. \label{eq:security}
\end{align}
\end{subequations}
Equation \eqnref{eq:logic} enforces commitment transitions; \eqnref{eq:uptime}--\eqnref{eq:dntime} impose minimum up- and downtimes; \eqnref{eq:pbound}--\eqnref{eq:headroom} define dispatch, ramping, and reserve headroom limits; and \eqnref{eq:balance}--\eqnref{eq:security} enforce network balance, DC line limits, and post-contingency security. Here, $UT_g$ and $DT_g$ are minimum up- and downtimes; $RU_g$ and $RD_g$ are ramp-up and ramp-down limits; $B_l$ is the line susceptance; $f_{l,t}^{(c)}$ is the flow on Line $l$ at Time $t$ after the outage of Contingency $c\in\cC$; and $F_l^{\max}$ and $F_{l,c}^{\max}$ are the base-case and post-contingency flow limits, respectively. As in standard market-grade SCUC formulations, the recovery layer keeps these physical constraints intact \cite{chen2023scuc,xavier2024ucjl}.

\subsection{Sequential Mapping and Reward}
We map partial commitment enforcement to a sequential decision process. At stage $t$, the agent observes
\begin{align}
\vect{s}_t=[\sin(2\pi t/T),\;\cos(2\pi t/T),\;\vect{u}_{t-1},\;\vect{z}_t],
\label{eq:state}
\end{align}
where the sinusoidal pair encodes the hour of day on the unit circle, so that temporally adjacent hours, including the midnight wrap-around, remain adjacent in feature space, as in standard cyclic positional encodings \cite{vaswani2017attention}; $\vect{u}_{t-1}\in\{0,1\}^{N_g}$ is the previous commitment vector; and $\vect{z}_t\in[0,1]^{N_g}$ is a quantum-sampled auxiliary feature vector (see \secref{sec:method}). The action is a multi-binary commitment proposal $\vect{a}_t\in\{0,1\}^{N_g}$ that is subsequently screened and partially enforced before the mixed-integer recovery solution.
The reward at each stage combines normalized recovery cost, switching, time, and infeasibility penalties with potential-based shaping, viz.,
\begin{align}
r_t^{\mathrm{RL}} &= -\Big(\lambda_c\tfrac{J_t}{\bar J}+\lambda_{\mathrm{sw}}\psi_t^{\mathrm{sw}}+\lambda_{\mathrm{time}}\psi_t^{\mathrm{time}}+\lambda_{\mathrm{inf}}\psi_t^{\mathrm{inf}}\Big) \nonumber\\
&\quad + \lambda_{\Phi}\big(\gamma\Phi_{t+1}-\Phi_t\big),
\label{eq:reward}
\end{align}
with the potential $\Phi_t = -\kappa\,\widehat C^{\mathrm{ED}}\big(\widehat D_{t+1}^{\mathrm{net}},\vect{u}_t\big)/\bar J$.
Here, $J_t$ is the recovered SCUC objective at Stage $t$; $\bar J$ is a cost normalizer; $\psi_t^{\mathrm{sw}}=N_g^{-1}\lVert\vect{a}_t-\vect{u}_{t-1}\rVert_1$ measures commitment switching; $\xi_t^{\mathrm{sol}}$ is the solver wall-clock time; and $\psi_t^{\mathrm{time}}=\log(1+\xi_t^{\mathrm{sol}}/\bar\xi^{\mathrm{sol}})$ penalizes slow recovery relative to the reference time, $\bar\xi^{\mathrm{sol}}$. The term $\psi_t^{\mathrm{inf}}$ aggregates load shedding, reserve shortfall, line flow, ramp, and minimum-time violations from the recovery layer. The nonnegative weights $\lambda_c$, $\lambda_{\mathrm{sw}}$, $\lambda_{\mathrm{time}}$, $\lambda_{\mathrm{inf}}$, and $\lambda_{\Phi}$ balance the cost, switching, timing, infeasibility, and shaping terms; the implementation uses $\lambda_c=1$, $\lambda_{\mathrm{sw}}=0.5$, $\lambda_{\mathrm{time}}=0.2$, and $\lambda_{\Phi}=0.5$, with the violation classes inside $\psi_t^{\mathrm{inf}}$ weighted between 1 (ramping) and 15 (load shedding); $\gamma\in(0,1)$ is the discount factor. In $\Phi_t$, $\widehat D_{t+1}^{\mathrm{net}}$ denotes the next-step net-load forecast; $\widehat C^{\mathrm{ED}}(\cdot)$ is a merit-order economic-dispatch surrogate; and $\kappa$ scales the shaping magnitude ($\kappa=1$). The reward is therefore grounded in the physically recovered schedule rather than in a purely learned proxy.

\section{Proposed HSAC-SCUC Method}
\label{sec:method}
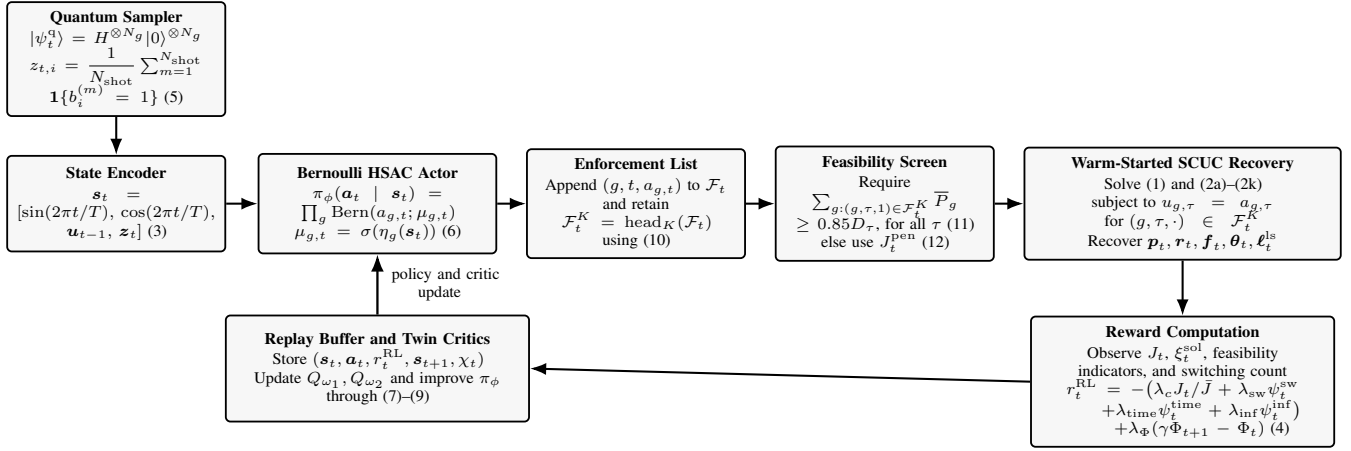
\begin{figure*}[t]
\centering
\resizebox{0.98\textwidth}{!}{\begin{tikzpicture}[
    font=\scriptsize,
    >=Latex,
    node distance=0.42cm and 0.42cm,
    box/.style={draw, rounded corners=2pt, line width=0.8pt, align=center, fill=black!3, inner sep=4pt},
    proc/.style={box, text width=3.05cm, minimum height=1.46cm},
    medproc/.style={box, text width=3.35cm, minimum height=1.52cm},
    wideproc/.style={box, text width=4.55cm, minimum height=1.70cm},
    diagproc/.style={box, text width=4.35cm, minimum height=1.62cm},
    arr/.style={-Latex, line width=0.9pt}
]

\node[proc] (state) {\textbf{State Encoder}\\[1pt]
$\vect{s}_t=[\sin(2\pi t/T),\,\cos(2\pi t/T),$\\
$\vect{u}_{t-1},\,\vect{z}_t]$ \eqnref{eq:state}};

\node[proc, above=0.58cm of state] (qfeat) {\textbf{Quantum Sampler}\\[1pt]
$|\psi_t^{\mathrm{q}}\rangle=H^{\otimes N_g}|0\rangle^{\otimes N_g}$\\
$z_{t,i}=\dfrac{1}{N_{\mathrm{shot}}}\sum_{m=1}^{N_{\mathrm{shot}}}$\\
$\mathbf{1}\{b_i^{(m)}=1\}$ \eqnref{eq:qfeat}};

\node[medproc, right=0.48cm of state] (actor) {\textbf{Bernoulli HSAC Actor}\\[1pt]
$\pi_\phi(\vect{a}_t\mid\vect{s}_t)=\prod_g \operatorname{Bern}(a_{g,t};\mu_{g,t})$\\
$\mu_{g,t}=\sigma\!\left(\eta_g(\vect{s}_t)\right)$ \eqnref{eq:policy}};

\node[proc, right=0.44cm of actor] (fix) {\textbf{Enforcement List}\\[1pt]
Append $(g,t,a_{g,t})$ to $\mathcal{F}_t$\\
and retain $\mathcal{F}_t^K=\operatorname{head}_K(\mathcal{F}_t)$\\
using \eqnref{eq:fixcap}};

\node[proc, right=0.44cm of fix] (screen) {\textbf{Feasibility Screen}\\[1pt]
Require $\sum_{g:(g,\tau,1)\in\mathcal{F}_t^K} \overline P_g$\\
$\ge 0.85D_\tau$, for all $\tau$ \eqnref{eq:filter}\\
else use $J_t^{\mathrm{pen}}$ \eqnref{eq:penalty}};

\node[wideproc, right=0.44cm of screen] (scuc) {\textbf{Warm-Started SCUC Recovery}\\[1pt]
Solve \eqnref{eq:obj} and \eqnref{eq:logic}--\eqnref{eq:security}\\
subject to $u_{g,\tau}=a_{g,\tau}$\\
for $(g,\tau,\cdot)\in\mathcal{F}_t^K$\\
Recover $\vect{p}_t,\vect{r}_t,\vect{f}_t,\vect{\theta}_t,\vect{\ell}_t^{\mathrm{ls}}$};

\node[diagproc, below=0.92cm of actor] (update) {\textbf{Replay Buffer and Twin Critics}\\[1pt]
Store $(\vect{s}_t,\vect{a}_t,r_t^{\mathrm{RL}},\vect{s}_{t+1},\chi_t)$\\
Update $Q_{\omega_1},Q_{\omega_2}$ and improve $\pi_\phi$\\
through \eqnref{eq:target}--\eqnref{eq:pi_loss}};

\node[diagproc, below=0.84cm of scuc] (reward) {\textbf{Reward Computation}\\[1pt]
Observe $J_t$, $\xi_t^{\mathrm{sol}}$, feasibility indicators, and switching count\\
$r_t^{\mathrm{RL}}=-\big(\lambda_cJ_t/\bar J+\lambda_{\mathrm{sw}}\psi_t^{\mathrm{sw}}$\\
$\qquad\ +\lambda_{\mathrm{time}}\psi_t^{\mathrm{time}}+\lambda_{\mathrm{inf}}\psi_t^{\mathrm{inf}}\big)$\\
$\qquad\ +\lambda_{\Phi}(\gamma\Phi_{t+1}-\Phi_t)$ \eqnref{eq:reward}};

\draw[arr] (qfeat) -- (state);
\draw[arr] (state) -- (actor);
\draw[arr] (actor) -- (fix);
\draw[arr] (fix) -- (screen);
\draw[arr] (screen) -- (scuc);
\draw[arr] (scuc) -- (reward);
\draw[arr] (reward.west) -- (update.east);
\draw[arr, line width=1.0pt] (update.north) -- node[right, inner sep=0.5pt, align=center] {\hspace{0.1cm} policy and critic\\update} (actor.south);

\end{tikzpicture}}
\caption{The proposed HSAC-SCUC workflow generates binary commitment proposals, retains at most $K$ chronological commitments for enforcement, rejects obviously undercommitted schedules through a capacity screen, and then recovers dispatch and security variables with the native SCUC model.}
\label{fig:framework}
\end{figure*}

\Figref{fig:framework} summarizes the module-level workflow, and \Algref{alg:hsac} gives the episode-level training loop.

\subsection{Quantum-Sampled State Augmentation}
The auxiliary channel follows the quantum feature-map paradigm, in which measurement outcomes of a quantum circuit supply features for a classical learner \cite{havlicek2019supervised}. The current implementation uses one qubit for each generating unit, prepared in the shallow reference state $|\psi_t^{\mathrm{q}}\rangle=H^{\otimes N_g}|0\rangle^{\otimes N_g}$, where $|0\rangle^{\otimes N_g}$ is the $N_g$-qubit all-zeros computational-basis state and $H^{\otimes N_g}$ applies a Hadamard gate to every qubit, producing the uniform superposition over all $2^{N_g}$ basis states; each qubit is then measured independently in the computational basis.\footnote{The quantum layer provides only stochastic side information; feasibility screening and SCUC recovery remain classical.} With $N_{\mathrm{shot}}$ measurement shots, letting $b_i^{(m)}\in\{0,1\}$ denote the measured value of Qubit $i$ in Shot $m$ and $\mathbf{1}\{\cdot\}$ the indicator function, the $i$th auxiliary feature is the empirical marginal
\begin{align}
z_{t,i}=\frac{1}{N_{\mathrm{shot}}}\sum_{m=1}^{N_{\mathrm{shot}}}\mathbf{1}\!\left\{b_i^{(m)}=1\right\}.
\label{eq:qfeat}
\end{align}
Quantum sampling is adopted because projective measurement natively returns multi-binary samples whose dimension matches the commitment vector, entering the state in \eqnref{eq:state} without a decoding layer, and because the channel establishes a hardware-compatible interface; replacing the Hadamard layer with a state-conditioned parameterized circuit turns the same interface into a trained quantum feature map whose output distributions are, in general, hard to sample classically \cite{havlicek2019supervised}. The corresponding limitation must be stated honestly; since the present circuit is state-independent, each $z_{t,i}$ is a scaled binomial draw with mean $1/2$ that a classical random-feature generator \cite{rahimi2007random} replicates exactly in distribution, so no quantum advantage is claimed, and the measured performance must be attributed to the HSAC-SCUC interface rather than to the quantum layer, which is retained as the integration point for the state-dependent circuits discussed in \secref{sec:conclusion}.

\subsection{Bernoulli HSAC for Multi-Binary Commitment}
SAC is adapted to multi-binary UC actions by factorizing the policy across generating units such that
\begin{align}
\pi_\phi(\vect{a}_t\mid\vect{s}_t)=\prod_{g=1}^{N_g}\operatorname{Bern}\!\left(a_{g,t};\mu_{g,t}\right),\quad \mu_{g,t}=\sigma\!\left(\eta_g(\vect{s}_t)\right).
\label{eq:policy}
\end{align}
Here, $\operatorname{Bern}(a;\mu)=\mu^{a}(1-\mu)^{1-a}$ is the Bernoulli probability mass function, $\eta_g(\vect{s}_t)$ is the actor-network logit for Unit $g$, and $\sigma(\cdot)$ is the logistic sigmoid; the actor thus outputs one Bernoulli mean for each unit. Two critics $Q_{\omega_1}$ and $Q_{\omega_2}$ mitigate overestimation. Using target critics $Q_{\bar\omega_1}$ and $Q_{\bar\omega_2}$, the temporal-difference target is
\begin{align}
y_t &= r_t^{\mathrm{RL}} + \gamma(1-\chi_t)\Big[\min_{j\in\{1,2\}} Q_{\bar\omega_j}(\vect{s}_{t+1},\tilde{\vect{a}}_{t+1}) \nonumber\\
&\hspace{2.0cm}-\alpha\,\mathcal{H}\big(\pi_\phi(\cdot\mid\vect{s}_{t+1})\big)\Big],
\label{eq:target}
\end{align}
where $\alpha$ is the entropy weight, $\chi_t\in\{0,1\}$ is the episode-termination indicator, $\tilde{\vect{a}}$ denotes an action generated by the current policy at the corresponding state, and $\mathcal{H}(\pi_\phi(\cdot\mid\vect{s}_t))$ is the sum of Bernoulli entropies across units. The critic and actor objectives, with expectations taken over minibatches drawn from the replay buffer $\mathcal{D}$, are
\begin{align}
L_Q &= \mathbb{E}\!\left[\left(Q_{\omega_j}(\vect{s}_t,\vect{a}_t)-y_t\right)^2\right], \label{eq:q_loss}\\
L_\pi &= -\mathbb{E}\!\left[\min_j Q_{\omega_j}(\vect{s}_t,\tilde{\vect{a}}_t)+\alpha\,\mathcal{H}\big(\pi_\phi(\cdot\mid\vect{s}_t)\big)\right].
\label{eq:pi_loss}
\end{align}
In the present implementation, the next-action target uses the Bernoulli mean, whereas the actor update uses sampled actions; this approximation avoids enumerating the $2^{N_g}$ action space while retaining entropy regularization \cite{haarnoja2018sac}.

\subsection{Feasibility-Aware Partial Commitment Enforcement}
The key interface between RL and SCUC is the accumulated enforcement list $\cF_t=\{(g,\tau,a_{g,\tau})\mid g\in\cG,\; \tau\le t\}$. The recovery model does not enforce every historical action. Instead, it retains only the first $K$-tuples,
\begin{align}
\cF_t^K = \firstK(\cF_t),\qquad K=20,
\label{eq:fixcap}
\end{align}
where the operator $\firstK(\cdot)$ keeps the $K$ earliest-appended tuples and discards the rest. The cap is fixed at $K=20$ across all systems so that the smallest fleets receive multiperiod guidance while the enforced block remains a small fraction of the $N_gT$ commitment binaries; \secref{sec:results} examines the consequences of this choice.
The active enforcement set is imposed through the equality constraints $u_{g,\tau}=a_{g,\tau}$ for all $(g,\tau,a_{g,\tau})\in\cF_t^K$; the resulting recovery problem preserves the original continuous and network-constrained physics, but it reduces the number of free commitment binaries to $N_{\mathrm{free}}(t)=N_gT-|\cF_t^K|$, i.e., when the cap is active, the unrestricted binary search space is reduced by a factor of $2^{|\cF_t^K|}$ relative to the unfixed commitment block. Before the mixed-integer solution, a capacity-based screen rejects clearly undercommitted schedules via
\begin{align}
\sum_{g:(g,\tau,1)\in \cF_t^K} \overline P_g \ge \beta D_\tau,\qquad \beta=0.85,\;\forall \tau\in\cT_t^K,
\label{eq:filter}
\end{align}
where $\cT_t^K\subseteq\cT$ is the set of periods that receive enforced tuples in $\cF_t^K$, and the threshold $\beta\in(0,1]$ sets the strictness of the screen; $\beta=0.85$ requires the enforced online units to cover at least 85\% of the period demand before the recovery solution is attempted. The screen is deliberately simple; since it ignores network, ramping, and reserve limits, it can pass schedules that recover only at a high cost or reject schedules that the solver could still complete economically. If \eqnref{eq:filter} fails, the environment returns a penalty objective
\begin{equation}
% \begin{aligned}
J_t^{\mathrm{pen}} = 10^6 + 10^3\sum_{\tau\in\cT}\Bigg[D_\tau - \sum_{g:(g,\tau,1)\in \cF_t^K} \overline P_g\Bigg]_+,
% \end{aligned}
\label{eq:penalty}
\end{equation}
where $[x]_+=\max(x,0)$.
A practically important consequence of the chronological rule in~\eqnref{eq:fixcap} is its limited temporal coverage; if each stage contributes $N_g$ new commitment variables, the number of fully covered periods and the fraction of the next covered period are
\begin{align}
N_{\mathrm{full}}=\left\lfloor\frac{K}{N_g}\right\rfloor,\qquad \sigma_{\mathrm{part}}=\frac{K-N_{\mathrm{full}}N_g}{N_g}.
\label{eq:coverage}
\end{align}
Hence, once $N_g>K$, the recovery model does not receive a fully enforced period; this coverage effect is central to the medium-scale results.

\begin{algorithm}[t]
\caption{HSAC-Guided SCUC with Partial Commitment Enforcement}
\label{alg:hsac}
\footnotesize
\begin{algorithmic}[1]
\State Initialize actor $\pi_\phi$, critics $Q_{\omega_1},Q_{\omega_2}$, target critics, replay buffer $\mathcal{D}$, and the warm-started SCUC model.
\For{episode $e=1,\ldots,N_{\mathrm{ep}}$}
    \State Reset the environment, obtain $\vect{u}_0$, and set $\cF\leftarrow\emptyset$.
    \For{$t=1,\ldots,T$}
        \State Construct $\vect{s}_t$ using \eqnref{eq:state} and \eqnref{eq:qfeat}; sample $\vect{a}_t\sim\pi_\phi(\cdot\mid\vect{s}_t)$.
        \State Append $(g,t,a_{g,t})$ to $\cF$ for all $g$ and keep $\cF_t^K$ using \eqnref{eq:fixcap}.
        \If{the capacity screen \eqnref{eq:filter} fails}
            \State Assign the penalty objective \eqnref{eq:penalty}.
        \Else
            \State Warm start the SCUC solver, enforce the tuples in $\cF_t^K$, and recover the remaining variables.
        \EndIf
        \State Compute $r_t^{\mathrm{RL}}$ from \eqnref{eq:reward}, store $(\vect{s}_t,\vect{a}_t,r_t^{\mathrm{RL}},\vect{s}_{t+1},\chi_t)$ in $\mathcal{D}$, and update the actor and critics using \eqnref{eq:target}--\eqnref{eq:pi_loss}.
    \EndFor
\EndFor
\end{algorithmic}
\end{algorithm}

\section{Implementation and Experimental Setup}
\label{sec:setup}
\Tabref{tab:impl} summarizes the fixed hyperparameters and solver settings shared by all experiments.
The actor and both critics use two hidden layers with 256 rectified linear unit (ReLU) activations. The replay buffer capacity is $2\times 10^5$, the batch size is 256, the discount factor is $\gamma=0.99$, the target-update rate (SAC target-network soft-update coefficient) is $\upsilon=5\times 10^{-3}$, and the entropy weight is $\alpha=0.05$. These learning hyperparameters follow standard SAC practice \cite{haarnoja2018sac} and were not tuned to the individual test systems; the influence of the interface parameters $K$ and $\beta$ is analyzed in \secref{sec:results}. The quantum auxiliary channel uses 128 shots per query with IBM Qiskit Aer 0.17.1 (SamplerV2 primitive). The recovery layer is implemented with UnitCommitment.jl, JuMP, and Gurobi \cite{xavier2024ucjl,lubin2023jump}; the test systems are the MATPOWER \mbox{14-,} \mbox{57-,} and 118-bus benchmark instances of UnitCommitment.jl, each with a 36-period horizon and its full contingency list. The implementation and episode traces are available in the companion repository \cite{hsacrepo2026}.
The evaluation uses representative traces and training runs of varying lengths for the three systems, spanning the practically relevant transition from $N_g<K$ to $N_g>K$. Since the recovery model is kept in the loop, each stated objective is measured after the SCUC recovery layer has enforced dispatch, reserve, and transmission constraints.

\begin{table}[t]
\caption{Key HSAC-SCUC Implementation Settings}
\label{tab:impl}
\centering
\footnotesize
\setlength{\tabcolsep}{4pt}
\begin{tabularx}{\columnwidth}{@{}lX@{}}
\toprule
\textbf{Component} & \textbf{Setting} \\
\midrule
Actor / critic & Two hidden layers, 256 ReLU units \\
Replay / batch & $2\times 10^5$ transitions / 256 \\
Discount / target update & $\gamma=0.99$, $\upsilon=5\times 10^{-3}$ \\
Entropy weight & $\alpha=0.05$ \\
Reward terms & Cost, switching, timing, infeasibility, shaping \\
Quantum channel & 128 shots/query; Aer 0.17.1 SamplerV2; Hadamard sampler \\
Recovery solver & UnitCommitment.jl + JuMP + Gurobi \\
Commitment enforcement & At most $K=20$ enforced binaries \\
Feasibility screen & Available committed capacity $\ge 0.85\times$ demand \\
Penalty objective & $10^6$ or $10^6 + 10^3\times$deficit \\
\bottomrule
\end{tabularx}
\end{table}

\section{Results and Discussion}
\label{sec:results}
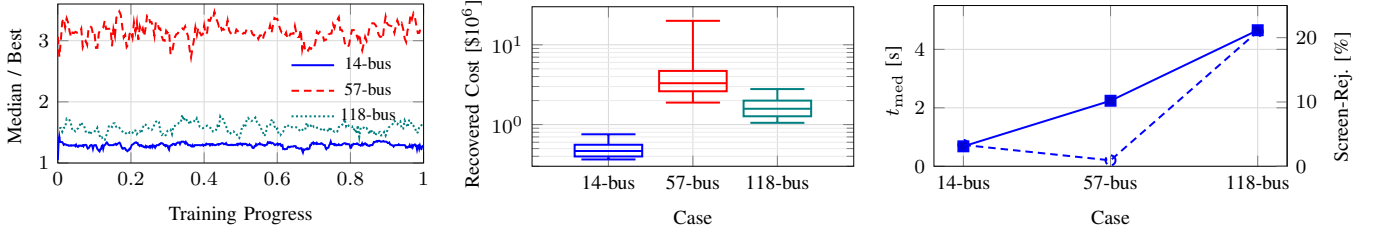
\begin{figure*}[t]
\centering
\hspace{-1.5cm}
\resizebox{0.91\textwidth}{!}{\begin{minipage}[t]{0.290\textwidth}
\centering
\begin{tikzpicture}
\begin{axis}[
    width=0.98\linewidth,
    height=0.426\textwidth,
    scale only axis,
    xmin=0, xmax=1,
    ymin=1.0, ymax=3.6,
    xlabel={Training Progress},
    ylabel={Median / Best},
    ylabel style={font=\footnotesize, xshift=0.2em, xshift=-0.3em},
    xlabel style={font=\footnotesize, yshift=-0.4ex},
    tick label style={font=\footnotesize},
    label style={font=\footnotesize},
    legend style={font=\scriptsize, draw=none, fill=none, at={(0.96,0.73)}, anchor=north east},
    grid=both,
    minor grid style={black!8},
    major grid style={black!15},
    every axis plot/.append style={thick},
    cycle list={blue, red, teal}
]
\addplot+[mark=none] table[x=Progress,y=RollingMedianNorm,col sep=comma] {figures/data/curve_14bus.csv};
\addlegendentry{14-bus}
\addplot+[mark=none, densely dashed] table[x=Progress,y=RollingMedianNorm,col sep=comma] {figures/data/curve_57bus.csv};
\addlegendentry{57-bus}
\addplot+[mark=none, densely dotted] table[x=Progress,y=RollingMedianNorm,col sep=comma] {figures/data/curve_118bus.csv};
\addlegendentry{118-bus}
\end{axis}
\end{tikzpicture}
\end{minipage}
% \hfill
\hspace{0.9cm}
\begin{minipage}[t]{0.255\textwidth}
\centering
\begin{tikzpicture}
\begin{axis}[
    width=0.98\linewidth,
    height=0.486\textwidth,
    scale only axis,
    ymode=log,
    ymin=0.3, ymax=30,
    ylabel={Recovered Cost [\$10$^6$]},
    ylabel style={font=\footnotesize, xshift=-0.3em, yshift=-0.3em},
    xlabel={Case},
    xlabel style={font=\footnotesize, yshift=-0.4ex},
    xtick={1,2,3},
    xticklabels={14-bus,57-bus,118-bus},
    x tick label style={font=\footnotesize},
    tick label style={font=\footnotesize},
    label style={font=\footnotesize},
    grid=both,
    minor grid style={black!8},
    major grid style={black!15},
    boxplot/draw direction=y,
    enlarge x limits=0.18,
    every axis plot/.append style={solid, thick},
    cycle list={blue, red, teal}
]
\addplot+[boxplot prepared={lower whisker=0.3651869549, lower quartile=0.3981803123, median=0.4661221126, upper quartile=0.5583810023, upper whisker=0.7550275385}] coordinates {};
\addplot+[boxplot prepared={lower whisker=1.8861925434, lower quartile=2.6151644126, median=3.2997451355, upper quartile=4.7041145480, upper whisker=19.9789617128}] coordinates {};
\addplot+[boxplot prepared={lower whisker=1.0516236341, lower quartile=1.2701408068, median=1.5782362812, upper quartile=1.9969178293, upper whisker=2.7860090115}] coordinates {};
\end{axis}
\end{tikzpicture}
\end{minipage}
% \hfill
\hspace{1.1cm}
\begin{minipage}[t]{0.280\textwidth}
\centering
\begin{tikzpicture}
\begin{axis}[
    width=0.98\linewidth,
    height=0.446\textwidth,
    scale only axis,
    symbolic x coords={14,57,118},
    xtick=data,
    xticklabels={14-bus,57-bus,118-bus},
    ymin=0, ymax=5.5,
    xlabel={Case},
    ylabel={$t_{\mathrm{med}}$ [s]},
    ylabel style={font=\footnotesize, xshift=0.2em, yshift=-0.2em},
    xlabel style={font=\footnotesize, yshift=-0.4ex},
    x tick label style={font=\footnotesize},
    tick label style={font=\footnotesize},
    label style={font=\footnotesize},
    grid=both,
    minor grid style={black!8},
    major grid style={black!15},
    every axis plot/.append style={thick}
]
\addplot+[mark=square*] coordinates {(14,0.6821) (57,2.2415) (118,4.6550)};
\end{axis}
\begin{axis}[
    width=0.98\linewidth,
    height=0.446\textwidth,
    scale only axis,
    symbolic x coords={14,57,118},
    xtick=data,
    axis y line*=right,
    axis x line=none,
    ymin=0, ymax=25,
    ylabel={Screen-Rej. [\%]},
    ylabel style={font=\footnotesize, xshift=-0.45em},
    tick label style={font=\footnotesize},
    label style={font=\footnotesize},
    every axis plot/.append style={thick}
]
\addplot+[mark=o, densely dashed] coordinates {(14,3.32) (57,0.92) (118,20.96)};
\end{axis}
\end{tikzpicture}
\end{minipage}}
\caption{Representative HSAC-SCUC experiments on the \mbox{14-,} \mbox{57-,} and 118-bus cases. \textit{Left}: rolling median (window of 2\% of the training episodes) over recovered episodes only, i.e., episodes that pass the capacity screen and complete the SCUC recovery solution. \textit{Center}: recovered final-cost distributions as 5/25/50/75/95 boxplots in \(\$10^6\). \textit{Right}: median episode solution time (solid line, squares, left axis) and capacity-screen rejection rate (dashed line, circles, right axis), i.e., the percentage of episodes assigned \(J_t^{\mathrm{pen}}\) since the enforced subset does not satisfy \eqnref{eq:filter}.}
\label{fig:results}
\end{figure*}

In a \textit{screen-rejected episode}, the capacity test in \eqnref{eq:filter} fails; the environment returns the penalty objective $J_t^{\mathrm{pen}}$ from \eqnref{eq:penalty} without invoking the SCUC recovery, and the episode is excluded from recovered-cost statistics. \Tabref{tab:summary} and \figref{fig:results} therefore separate the quality of the recovered SCUC objective from the frequency with which the enforced subset is rejected before recovery.

\begin{table}[h]
\caption{Representative HSAC-SCUC Runs and Recovered Costs in \(\$10^6\)}
\label{tab:summary}
\centering
\footnotesize
\setlength{\tabcolsep}{3pt}
\begin{tabular}{lccccccc}
\toprule
\textbf{System} & $N_g$ & \textbf{Ep.} & $J_{\mathrm{best}}^{\mathrm{rec}}$ & $J_{50}^{\mathrm{rec}}$ & $J_{90}^{\mathrm{rec}}$ & $t_{\mathrm{med}}$ & \textbf{Screen-Rej.}\% \\
\midrule
14-bus & 5 & 10{,}000 & 0.361 & 0.466 & 0.673 & 0.68 & 3.32 \\
57-bus & 7 & 5{,}000 & 1.049 & 3.300 & 16.234 & 2.24 & 0.92 \\
118-bus & 54 & 5{,}000 & 1.000 & 1.578 & 2.449 & 4.66 & 20.96 \\
\bottomrule
\end{tabular}
\end{table}
The 14-bus case remains the strongest result (see \tabref{tab:summary}). The left panel of \figref{fig:results} shows that the rolling recovered-cost median settles near $1.30\times$ the best recovered cost, indicating that the SCUC solver repeatedly receives useful commitment restrictions; the right panel shows that this behavior is achieved with a subsecond median episode time and a 3.32\% rejection rate.
The 57- and 118-bus cases reveal two distinct degradation modes. In the 57-bus case, the capacity-screen rejection rate remains below 1\%, indicating that the policy has learned commitment patterns that are usually compatible with the fixed startup, shutdown, reserve, and network constraints enforced in the recovery layer. However, the distribution of recovered costs is wide (see \tabref{tab:summary}); the actor often proposes schedules that survive screening but remain expensive after the SCUC model reconstructs dispatch, reserves, and line-feasible flows. The larger cost values in the 57-bus case should not be interpreted as a direct statement that it is harder than the 118-bus case.\footnote{Absolute dollar magnitudes are system-specific (different load levels, fleets, and cost coefficients), so cross-system costs should not be compared directly.} In the 118-bus case, by contrast, the rejection rate rises to 20.96\%, but the successful recoveries are comparatively tighter (see \tabref{tab:summary}), and about 79\% of episodes still recover successfully. In other words, the 57-bus case is dominated by expensive feasible recoveries, whereas the 118-bus case is dominated by undercoverage of the binary interface itself.
For an unassisted reference point, we solved the full-horizon instances directly, with no policy guidance, using the same UnitCommitment.jl and Gurobi stack and the solver settings of the training loop; on a desktop workstation, the solutions require approximately 0.1\,s, 0.2\,s, and 1.9\,s of wall-clock time for the \mbox{14-,} \mbox{57-,} and 118-bus instances with the 2017-01-01 daily profiles, and the resulting 14-bus optimal cost of \(\$3.606\times 10^5\) coincides with the best recovered objective in \tabref{tab:summary}.\footnote{The episode times in \tabref{tab:summary} were measured on a different machine and include policy inference, quantum sampling, and all stage-wise recovery solutions of an episode, so the timing comparison is indicative rather than strictly controlled.} Therefore, the interface transmits enough information for the recovery model to attain the full-horizon optimum of the smallest case, but the exact solver is not the computational bottleneck at these benchmark sizes; the study should be read as a controlled analysis of how much commitment information a policy can usefully transmit to a market-grade solver, with wall-clock benefits possible only at scales where the unassisted MILP itself becomes limiting.
These results clarify both the novelty and the limitations of the proposed approach. Unlike end-to-end RL UC, the policy need not satisfy network, reserve, or contingency constraints, and unlike quantum UC based on QAOA, annealing, or QUBO encodings, the method never optimizes the combinatorial block on quantum hardware; the quantum component is only a state-side auxiliary channel. The actual novelty lies in the solver-preserving interface whereby RL proposes commitment binaries, the recovery model enforces only the first $K$ chronological tuples, and the original SCUC solver certifies feasibility within the restricted search space. This positioning is, in spirit, close to recent solver-compatible commitment-reduction methods \cite{wang2026structure,xiong2025successive}, but the mechanism here is interaction-driven RL rather than supervised masking or linear-programming (LP)-guided restriction.

\begin{figure}[t]
\centering
\resizebox{0.98\columnwidth}{!}{\begin{tikzpicture}
\begin{groupplot}[
    group style={group size=1 by 2, vertical sep=0.72cm},
    width=0.94\columnwidth,
    height=0.42\columnwidth,
    xmode=log,
    log basis x=10,
    xmin=200, xmax=12000,
    grid=both,
    minor grid style={black!8},
    major grid style={black!15},
    tick label style={font=\footnotesize},
    label style={font=\footnotesize},
    every axis plot/.append style={thick},
    cycle list={blue, red, teal}
]
\nextgroupplot[
    ylabel={Median Cost [\$10$^6$]},
    ymin=0.3, ymax=3.7,
    xticklabels=\empty,
    legend style={font=\scriptsize, draw=none, fill=none, at={(0.98,0.90)}, anchor=north east}
]
\addplot+[mark=*] table[x=Episodes,y=MedianCostMillion,col sep=comma] {figures/data/budget_14bus.csv};
\addlegendentry{14-bus case}
\addplot+[mark=square*, densely dashed] table[x=Episodes,y=MedianCostMillion,col sep=comma] {figures/data/budget_57bus.csv};
\addlegendentry{57-bus case}
\addplot+[mark=triangle*, densely dotted] table[x=Episodes,y=MedianCostMillion,col sep=comma] {figures/data/budget_118bus.csv};
\addlegendentry{118-bus case}

\nextgroupplot[
    ylabel={Rejection Rate [\%]},
    xlabel={Episodes},
    xlabel style={yshift=-0.4ex},
    ymin=0, ymax=25
]
\addplot+[mark=*] table[x=Episodes,y=FallbackPct,col sep=comma] {figures/data/budget_14bus.csv};
\addplot+[mark=square*, densely dashed] table[x=Episodes,y=FallbackPct,col sep=comma] {figures/data/budget_57bus.csv};
\addplot+[mark=triangle*, densely dotted] table[x=Episodes,y=FallbackPct,col sep=comma] {figures/data/budget_118bus.csv};
\end{groupplot}
\end{tikzpicture}}
\caption{Sensitivity to the number of training episodes for the \mbox{14-,} \mbox{57-,} and 118-bus cases. The upper panel shows the median episode objective over all episodes, in \(\$10^6\), whereas the lower panel shows the capacity-screen rejection rate.}
\label{fig:budget}
\end{figure}
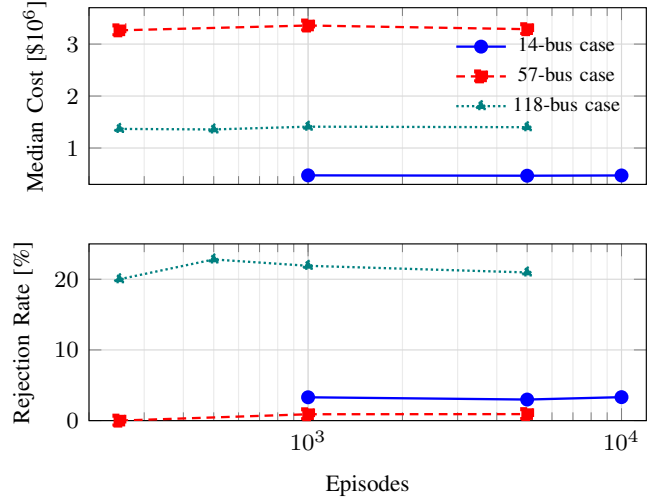

\Figref{fig:budget} shows that longer training runs do not by themselves remove the dominant error mechanisms; across the evaluated training lengths, the median episode objective of every system stays within a narrow band (e.g., \(\$1.36\)--\(\$1.40\times 10^6\) on the 118-bus case between 250 and 5{,}000 episodes), and the rejection rates barely move. Hence, although the larger systems were trained with fewer episodes than the 14-bus case (see \tabref{tab:summary}), the flat trends indicate that the bottleneck is not simply an insufficient number of RL iterations. The screening threshold interacts with the training length; raising $\beta$ above 0.85 converts additional proposals into screen-rejected episodes and thus withholds recovered-cost feedback from the critic precisely where it is already scarce, whereas lowering $\beta$ admits more undercovered proposals whose recovered cost is dominated by solver-completed commitments, so the recovered cost no longer reflects the policy's own decisions. Varying $\beta$ therefore mainly shifts episodes between the rejection-count and recovered-cost tails; more successful 118-bus recoveries are expected from coverage-aware enforcement, as discussed below, rather than from retuning the screen alone.
A more informative explanation comes from the coverage analysis implied by \eqnref{eq:coverage}. The enforcement-regime column of \tabref{tab:diag} shows that the chronological enforcement rule rapidly loses temporal reach as $N_g$ grows; with $K=20$, the 14-bus case fully covers four periods ($K/N_g=4.0$), the 57-bus case covers two full periods plus 0.86 of the third ($K/N_g\approx 2.86$), and once $N_g>K$, as in the 118-bus case ($K/N_g\approx 0.37$), the recovery model no longer receives even one complete period of commitment enforcement.

The deterioration in the 118-bus case is therefore coupled to the combinatorial information that the recovery model receives; once the number of units exceeds the cap, the signal passed from the policy to the optimizer becomes very small, and the present subset-selection rule is too coarse for medium-scale SCUC.

The temporal asymmetry of the enforcement rule also matters. Since tuples are appended in period order, the enforced subset always concentrates on the earliest periods of the horizon, whereas minimum up- and downtimes, ramp constraints, reserve trajectories, and contingency-feasible dispatch are all temporally coupled. If only 37\% of period 1 is enforced, as in the 118-bus case, the recovery solver reconstructs nearly the entire commitment trajectory on its own.

\begin{table}[h]
\caption{Coverage and Policy Characteristics under \(K=20\)}
\label{tab:diag}
\centering
\footnotesize
\setlength{\tabcolsep}{3pt}
\begin{tabularx}{\columnwidth}{@{}lcccX@{}}
\toprule
\textbf{System} & \textbf{On-fraction} & \textbf{Avg. stages} & \textbf{Median shots} & \textbf{Enforcement regime} \\
\midrule
14-bus & 0.501 & 3.98 & 384 & four full periods \\
57-bus & 0.500 & 31.37 & 4{,}608 & two full periods + 0.86 of period 3 \\
118-bus & 0.500 & 36.00 & 4{,}608 & 0.37 of period 1 \\
\bottomrule
\end{tabularx}
\end{table}

\Tabref{tab:diag} reinforces the same conclusion from a different angle; here, \textit{on-fraction} denotes the average share of units with $a_{g,t}=1$, \textit{average stages} denotes the mean number of sequential decision stages completed before termination, and \textit{median shots} denotes the median number of measurement shots per episode. The on-fraction remains essentially 0.5 across all three cases, indicating that the current Bernoulli actor remains highly exploratory. This exploratory behavior preserves feasibility discovery but also slows convergence toward more selective, low-cost subsets as the action space grows. In entropy terms, an on-fraction of 0.5 means that the factorized policy in \eqnref{eq:policy} stays near its maximum entropy of $N_g\ln 2$, so the entropy-regularized objective with $\alpha=0.05$ continues to favor exploration at these training lengths; the rolling medians in the left panel of \figref{fig:results} quantify the resulting slow decrease of recovered cost, and the underlying cost, action, and timing traces are available in the repository \cite{hsacrepo2026}. The difference across cases is therefore driven not by a different sparsity pattern but by how much of the sampled binary information the recovery model is allowed to use.

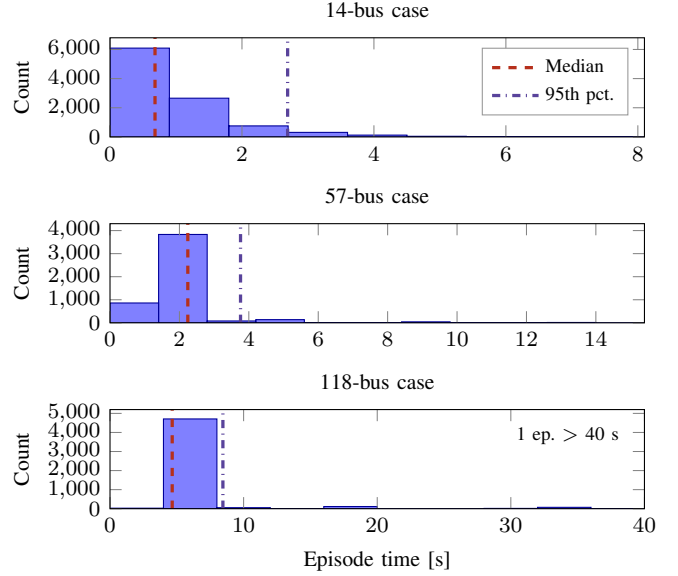
\begin{figure}[t]
\centering
\resizebox{\columnwidth}{!}
{% Per-case histograms of training time per episode, rebuilt natively in pgfplots
% from figures/data/hist_{14,57,118}bus.csv and runtime_quantiles.csv so that the
% median and 95th-percentile markers use distinct colors and dash patterns.
\resizebox{0.98\columnwidth}{!}{%
\begin{tikzpicture}
\begin{groupplot}[
  group style={group size=1 by 3, vertical sep=1.12cm},
  scale only axis,
  width=6.9cm,
  height=1.28cm,
  ymin=0,
  ylabel={Count},
  ylabel near ticks,
  scaled y ticks=false,
  tick label style={font=\footnotesize},
  label style={font=\footnotesize},
  title style={font=\footnotesize, yshift=-2pt},
  legend style={font=\scriptsize, fill=white, draw=black!40, at={(0.97,0.90)}, anchor=north east, legend cell align=left},
  axis on top,
]
% ---------------- 14-bus ----------------
\nextgroupplot[title={14-bus case}, xmin=0, xmax=8.1, ymax=6800,
  ytick={0,2000,4000,6000}]
\addplot [ybar interval, forget plot, fill=blue!50, draw=blue!60!black] coordinates {
(0.0,6091) (0.9,2656) (1.8,756) (2.7,311) (3.6,124) (4.5,43) (5.4,10) (6.3,7) (7.2,2) (8.1,0)
};
\addplot [BrickRed, dashed, very thick] coordinates {(0.682,0) (0.682,6800)};
\addlegendentry{Median}
\addplot [Violet, dashdotted, very thick] coordinates {(2.692,0) (2.692,6800)};
\addlegendentry{95th pct.}
% ---------------- 57-bus ----------------
\nextgroupplot[title={57-bus case}, xmin=0, xmax=15.4, ymax=4300,
  ytick={0,1000,2000,3000,4000}]
\addplot [ybar interval, forget plot, fill=blue!50, draw=blue!60!black] coordinates {
(0.0,860) (1.4,3832) (2.8,80) (4.2,138) (5.6,7) (7.0,6) (8.4,42) (9.8,9) (11.2,11) (12.6,15) (14.0,0) (15.4,0)
};
\addplot [BrickRed, dashed, very thick] coordinates {(2.242,0) (2.242,4300)};
\addplot [Violet, dashdotted, very thick] coordinates {(3.761,0) (3.761,4300)};
% ---------------- 118-bus ----------------
\nextgroupplot[title={118-bus case}, xmin=0, xmax=40, ymax=5200,
  ytick={0,1000,2000,3000,4000,5000},
  xlabel={Episode time [s]}]
\addplot [ybar interval, forget plot, fill=blue!50, draw=blue!60!black] coordinates {
(0.0,27) (4.0,4708) (8.0,54) (12.0,6) (16.0,114) (20.0,0) (24.0,0) (28.0,16) (32.0,72) (36.0,2) (40.0,0)
};
\addplot [BrickRed, dashed, very thick] coordinates {(4.655,0) (4.655,5200)};
\addplot [Violet, dashdotted, very thick] coordinates {(8.449,0) (8.449,5200)};
\node [font=\scriptsize, anchor=north east] at (rel axis cs:0.97,0.92) {1 ep. $>$ 40 s};
\end{groupplot}
\end{tikzpicture}%
}}
\caption{Histograms of training time per episode from the raw summary traces. The dashed red and dash-dotted violet vertical lines mark the median and the 95th percentile, respectively; one episode above 40 s in the 118-bus case is omitted for readability.}
\label{fig:runtime}
\end{figure}

\Figref{fig:runtime} completes the discussion with the distribution of training time per episode. The histogram for the 14-bus case is broad relative to its median since some episodes recover quickly, whereas others incur additional branch-and-bound work. The 57-bus histogram is sharply centered around 2--3 s, consistent with frequent recovery solutions that are usually feasible but sometimes expensive. The 118-bus histogram concentrates near 4.5--5 s but also exhibits a small secondary cluster in the right tail; most recovery solutions behave repeatably once the screen is passed, whereas a small subset incurs substantially higher mixed-integer effort. The free-binary count $N_{\mathrm{free}}$ provides another useful interpretation; for a fixed cap $K$, the recovery layer always removes the same number of binary variables, so the relative impact of the reduction shrinks rapidly as the commitment block grows, and the meaningful multiperiod reduction seen on the 14-bus case becomes only a weak perturbation of the much larger 118-bus binary polytope. The cap value itself is a compromise; lowering $K$ drives the recovery model toward the unassisted MILP and the guidance signal vanishes, whereas raising $K$ extends temporal coverage and enlarges the $2^{|\cF_t^K|}$ search-space reduction, but every enforced binary is an equality constraint chosen by an exploratory actor, so an aggressive cap increases both the rejection frequency and the risk of expensive recoveries. Covering one full period of the 118-bus case already requires $K\ge 54$, and proportional multiperiod guidance requires the cap to grow with the fleet size (e.g., $K=2N_g$); the coverage ratio $K/N_g$, rather than the absolute cap, is the operative design quantity. A stronger medium-scale design should therefore adapt $K$ to system size or to rank candidate commitment restrictions based on operational sensitivity, e.g., ramp-critical, reserve-critical, or congestion-sensitive units.

\section{Conclusion}
\label{sec:conclusion}
This paper presented a three-layer hybrid framework for SCUC in which RL proposes commitment binaries, a quantum-sampled auxiliary channel enriches the state, and a native mixed-integer SCUC model enforces dispatch and security feasibility after partial commitment. The method preserves the original SCUC model, keeping the learning interface transparent, modular, and compatible with market-grade optimization software.
The 14-bus case yields stable low-cost recovery, attaining the full-horizon optimum, with a sub-second median episode time and a low screen-rejection rate; the 57-bus case invokes the recovery model in almost every episode, yet its recovered-cost distribution remains heavy-tailed; and the 118-bus case, where $N_g>K$, loses complete-period coverage, and its rejection rate rises sharply. The key technical result is the coverage analysis; with a fixed cap of $K=20$, the commitment information passed to the solver collapses once the number of units exceeds the cap. Future work will replace the shallow quantum sampler with parameterized quantum circuits (PQCs) and pair them with adaptive or confidence-ranked commitment enforcement on larger test systems. Benchmarking against stronger classical baselines, e.g., a classical random-feature control for the auxiliary channel trained under identical settings \cite{rahimi2007random}, learning-based branching, structural masking, successive restriction, and hybrid acceleration \cite{dai2025reduction,wang2026structure,xiong2025successive,barrass2025leveraging}, will clarify how much of the remaining limitation is algorithmic and how much stems from the fixed-cap interface.

\section*{Acknowledgment}
The computation for this work was performed on the University of Missouri's Quantum Innovation Center, in partnership with IBM Quantum and facilitated by Research Support Services at the University of Missouri, Columbia, MO. DOI: 10.32469/10355/107781

\balance
\bibliographystyle{IEEEtran}
\bibliography{References}

\end{document}